\documentclass[a4paper]{article}
\pdfoutput=1
\usepackage{icrc2013}
\usepackage{subfigure}
\usepackage[english]{babel}

\title{Geomagnetic Backtracing: A comparison of Tsyganenko 1996 and 2005 External Field
models with AMS-02 data}

\shorttitle{Geomagnetic Backtracing and AMS-02}

\authors{M. J. Boschini$^{1,2}$, C. Consolandi$^{1,\dag}$, S. Della Torre$^{1}$, M. Gervasi$^{1,3}$, D. Grandi$^{1}$, S. Haino$^{4}$, G. La Vacca$^{1,3}$, S. Pensotti$^{1,3}$, P.G. Rancoita$^{1}$, D. Rozza$^{1,5,6}$, M. Tacconi$^{1,3}$}

\afiliations{
$^1$ \textit{INFN Sezione di Milano Bicocca, I-20126 Milano, Italy} \\ 
$^2$ \textit{CINECA, I-20090 Segrate (MI), Italy} \\
$^3$ \textit{Universit\`a di Milano Bicocca, I-20126 Milano, Italy} \\ 
$^4$ \textit{National Central University, NCU, Chung-Li, Tao Yuan 32054, Taiwan} \\ 
$^5$ \textit{Universit\`a dell'Insubria, I-22100 Como, Italy} \\ 
$^6$ \textit{European Organization for Nuclear Research, CERN, CH-1211 Geneva 23, Switzerland} \\
\scriptsize{
$^{\dag}$ now at: University of Hawaii at Manoa, Hi-96822 Honolulu, USA.
}
}

\email{Davide.Rozza@mib.infn.it}

\abstract{
We used a backtracing code to reconstruct particle trajectory inside the Earth Magnetosphere during the last 
solar active period (2011 and 2012) when very high Solar Wind pressure values were measured.
We compared our results on AMS-02 proton and electron data with 2 different External Field models, namely
Tsyganenko 1996 (T96) and 2005 (T05), both for quiet (defined as the periods when the solar wind pressure is 
below the average value, set at 2nPa) and active periods. Although T05 has been specifically 
designed for storm events, at high energy the particle trajectory is similar for the two models.
For instance at rigidities larger than 50 GV, the RMS of angular difference between reconstructed asymptotic 
direction outside the Magnetosphere is of the order of the millirad, while it increases 
at intermediate energies.
We also confirmed, as a function of the pointing direction, the well known East-West effect on the trajectory 
of primary particles and on the access solid angle on the AMS detector. 
}


\begin{document}
\maketitle

\section{Introduction}
To study the effect of the Earth magnetosphere on cosmic rays detected by the
AMS experiment during its mission on board of the ISS (sinche May 2011) 
we developed a software code for particle tracing. 
Particles trajectories have been evaluated in the framework of the
internal magnetic field IGRF-11 (\cite{bib:IGRF}) and the external magnetic field Tsyganenko
96 and 2005 (\cite{bib:T96} \cite{bib:T05}). Several effects of the Earth magnetic field on charged particles outside
the atmosphere have been studied.
Our attention was focused on the two external field models, the first, Tsyganenko 96 (hereafter T96) 
that was developed for quiet geomagnetic periods, and the second, Tsyganenko 2005 (hereafter T05) that was developed with
new data from 1996 to 2000, especially for storm events.
For this study we were able to use values of the solar
parameters measured during 2011 and 2012. Parameters for T96 were obtained from OMNIWEB (\cite{bib:omni}) 
while parameters for T05 from the web repository \cite{bib:05REP} or directly provided by N. Tsyganenko itself.
The backtracing provides the separation of allowed and forbidden 
primary particle trajectories and can be used to distinguish between primary and secondary CR.
Among the Galactic Cosmic Rays (GCR), protons are largely the most
abundant, but also the amount of Helium nuclei and electrons is
relevant. In this work we selected only AMS-02 electrons and, in order to evaluate the effect of the external field, 
our analysis was restricted to primary (allowed) particles.
Our results can be used for the study on electron and positron anisotropy, 
in oder to take in proper account the effect of the Earth magnetic field, 
even at energies around 100 GeV. 
The presence of antimatter in cosmic rays (mostly positrons), has been observed by the AMS-02 experiment
\cite{bib:phys-rev}. The fraction of positrons resulted to be increasing with energy, up to 350 GeV, putting the origin of this signal.  
Local astrophysical sources  (within 1 kpc) like pulsars are not yet confirmed 
as possible explanation (see \cite{bib:ROZZA}). Further information can be obtained looking for the anisotropy 
in the spatial distribution of these particles. 

\section{Particle Tracing in the Geomagnetic Field}
The Earth magnetic field can be roughly seen as a magnetic dipole, whose axis
is shifted from the Earth center of \( \sim 500\: km \), inclined of \( \sim 11^{\circ } \)
and opposite to the geographic rotational axis. Because of the presence of the
solar wind, the magnetic field is hardly compressed in the dayside and smootly
decompressed in the nightside, creating a highly asymmetric magnetosphere configuration.
Inside this region, due to the Lorentz force, positive charged particles
coming from outside the magnetosphere are rotated clockwise and negative charged
particles counterclockwise.
To describe the effect of magnetic fields on charged particles we can use the
magnetic rigidity \( R \), essentially the ratio between the momentum of the
particle and its charge (\( R=\frac{pc}{Ze} \)). 
Particle trajectory in the ``almost dipolar'' Earth magnetic field is more curved 
the lower is its momentum (energy), so  from previous 
formula we can define
for every point in space a limit called \emph{rigidity cut-off} \cite{bib:fermi},
below which primary cosmic rays will never reach any detector in orbit around
the Earth. 
\begin{equation}
R_{cut}\geq 59.6\cdot \left[ \frac{1-\sqrt{1-\cos ^{3}\vartheta _{m}\cdot \cos \gamma }}{\cos \vartheta _{m}\cdot \cos \gamma }\right] ^{2}
\end{equation}
Where $R_{cut}$ is in $GV$, $\vartheta _{m}$ is the magnetic
latitude (see par.4) and $\gamma$ is the angle between particle velocity
and East-West geomagnetic direction (from East to West).

We need to reconstruct the particle trajectory to study the properties of cosmic
rays detected by AMS, especially across the rigidity cut-off. Eq.1 can still be useful only in relation with an empirical 
estimate of the confidence level of this sharp edge. We define $R_{cut}$ from the particle tracing: 
if a particle trajectory comes from the external border of the magnetosphere, its rigidity is
above the cut-off, if comes from the Earth, it is below. This can be done using
a code that reproduces the interaction of a charged particle with all the
magnetic fields along the whole path.

The study of geomagnetic field and its interaction with charged particles in
orbit around the Earth is developing very fast. Mainly due to recent data about
intramagnetospheric magnetic fields and the sun interaction with the geomagnetic
dipole new opportunities of studying the geomagnetic effects are provided.

We developed a code using internal magnetic field model IGRF 11 (\cite{bib:IGRF}),
and the latest external magnetic field models Tsyganenko 1996 and 2005(\cite{bib:tsygan}). 
This program (see \cite{bib:ams_mib})
has been optimized and adapted in order to analise the data of the AMS experiment
(see \cite{bib:ams}), a magnetic spectrometer for cosmic rays that has been installed on the
International Space Station in May 2011 during STS-134 Shuttle flight.

With this analysis it has been possible to investigate the composition of secondary
and primary cosmic rays in the overall measured spectrum as a function of local
geomagnetic and geographic positions. Our attention was especially focused
on primary CR and the accuracy of incoming direction reconstructed outside of the
magnetosphere for anisotropy study (see \cite{bib:anys}).

\subsection{Internal and External Field Models}
\label{subsec:int_ext}
The internal magnetic fied model, IGRF 11, is a mathematical description of the Earth main magnetic field and its annual rate of change (secular variation).
In source free regions this main field is the negative gradient of a scalar potential $V$, represented by a truncated series of spherical
(13$^{th}$  order ) harmonic expansion (\cite{bib:igrfweb}):
\begin{eqnarray}
 V(R,\vartheta,\lambda) =  R_{E}\cdot\sum_{n+1}^{N}\sum_{m=0}^{n}\left(\frac{R_{E}}{r}\right)^{n+1} \nonumber \\
	    \cdot [g^{m}_{n}\cos(m\phi)+ h^{m}_{n}\sin(m\phi)] \nonumber \\
	    \cdot P^{m}_{n}\cos(\vartheta)
\end{eqnarray}
The so-called external magnetic field models, mainly developed by N. Tsyganenko, describes all the magnetic contributions originated
outside the Earth surface. These can be charge particle currents, like the Ring current
or magnetic interconnection fields, for example at the border of the magnetopause where the geomagnetic fields is shielding us from the
Solar Wind.

The last and most updated Tsyganenko models are T96 \cite{bib:refT96} and T05 \cite{bib:refT05}. The first has been developed with a new set of
measurements performed by different satellites, and includes all currents insite the magnetosphere (Birkeland regions, ring currents, tail
currents, interconnection field and magnetopause field) in addition with a solar wind controlled magnetopause based on Sibeck equation
(\cite{bib:sibeck}) with a compression/decompression factor depending on solar wind ram pressure. 
This model depends on 4 parameters measured by a network of satellites and that can be retrieved thanks to NASA OMNIWEB service
(\cite{bib:omni}), these parameters are: 
the RAM pressure of the Solar Wind $P_{dyn}$, the Disturbance Sorm Time index $Dst$ (\cite{bib:dst}), and the $Y$ and $Z$ components of the 
Interplanetary Magnetic Field $B_{Y}^{IMF}$ and $B_{Z}^{IMF}$.
T05 has been developed with a new database of measurements, from 1996 to 2000, and has been explicitly designed for storm events (it has
been fitted on 37 major events during this period) introducing also an asymmetric time scale 
of the magnetosphere disturbancies due to solar events. All components included in T96 are also present here even if they depend on
additiona parameters and include additonal components to reproduce measured data (for example the tail includes a fast and a slow
component). These new 6 parameters, called $W1$ ... $W6$  have been fitted during the 37 major events of the 1996-200 period and depend only from 3 measured
parameters as the Solar Wind density $N_{sw}$ and speed $V_{sw}$ , and the southward inteplanetary magnetic field $B_{}$.
We were able to obtain the external field parameters (for T96 they are) on a 5 minutes time base but for 
the AMS data analysis due to the large variations, especially for the Dst index, 
we choosed to use as first step approximation
the average hourly values of the four parameters (for T96) and the additional 6 parameters (for T05)
for the whole period of the mission (Starting from May 2011 up top the end of 2012). 

\section{External or Not External}
The CPU time (\cite{bib:ams_mib}) needed to estimate the external field with respect to the inner part is dominant (it depends on the particle 
position and direction but sometime can also be 10 times bigger), and there are differences even between the two external models, so T05
is almost 2 times slower than T96. 
The first question to be solved so is if an external field is needed. A macroscopic difference is that the IGRF
representation of the geomagnetic field is essentially symmetric, while (\cite{bib:tsygan})
the magnetosphere is highly asymmetric. This consideration can suggest the introduction of the external magnetic field. 
We anyway tested our backtracing code in different situations: excluding the external field (only internal one) or using
both T05 and T96 models. As shown in Table \ref{table_noext}, we backtraced a sample of $2.5\times10^{6}$
simulated protons between 0.3 and 200 $GV$ at the ISS altitude and uniformly inside the AMS-02 field of view (45$^{\circ}$ from the zenith). 
Almost 20\% of them are recognized as primary CR (allowed trajectories) only when traced back with the internal 
field, while they are not more primary CR when traced back with the complete (internal plus external) magnetic field. 
Most of this difference is located at high latitudes where the 
difference is close to 100\%. Changing the external field model the difference is below 10\%.

\begin{table}[ht!]
\begin{center}
\begin{tabular}{|c|c|c|c|}
\hline Model & IGRF & IGRF + T96 & IGRF + T05 \\ \hline
IGRF   & 0\% & 21.5\% & 19.8\%\\ \hline
IGRF + T96   &  & 0\% & 9.5\%\\ \hline
IGRF +  T05 &  &  & 0\%\\ \hline
\end{tabular}
\caption{Difference nature of particles - allowed or forbidden trajectoy - with different magnetic field models}
\label{table_noext}
\end{center}
\end{table}

\begin{figure}[ht!]
\hfill
\subfigure[Latitude]{\includegraphics[width=0.235\textwidth]{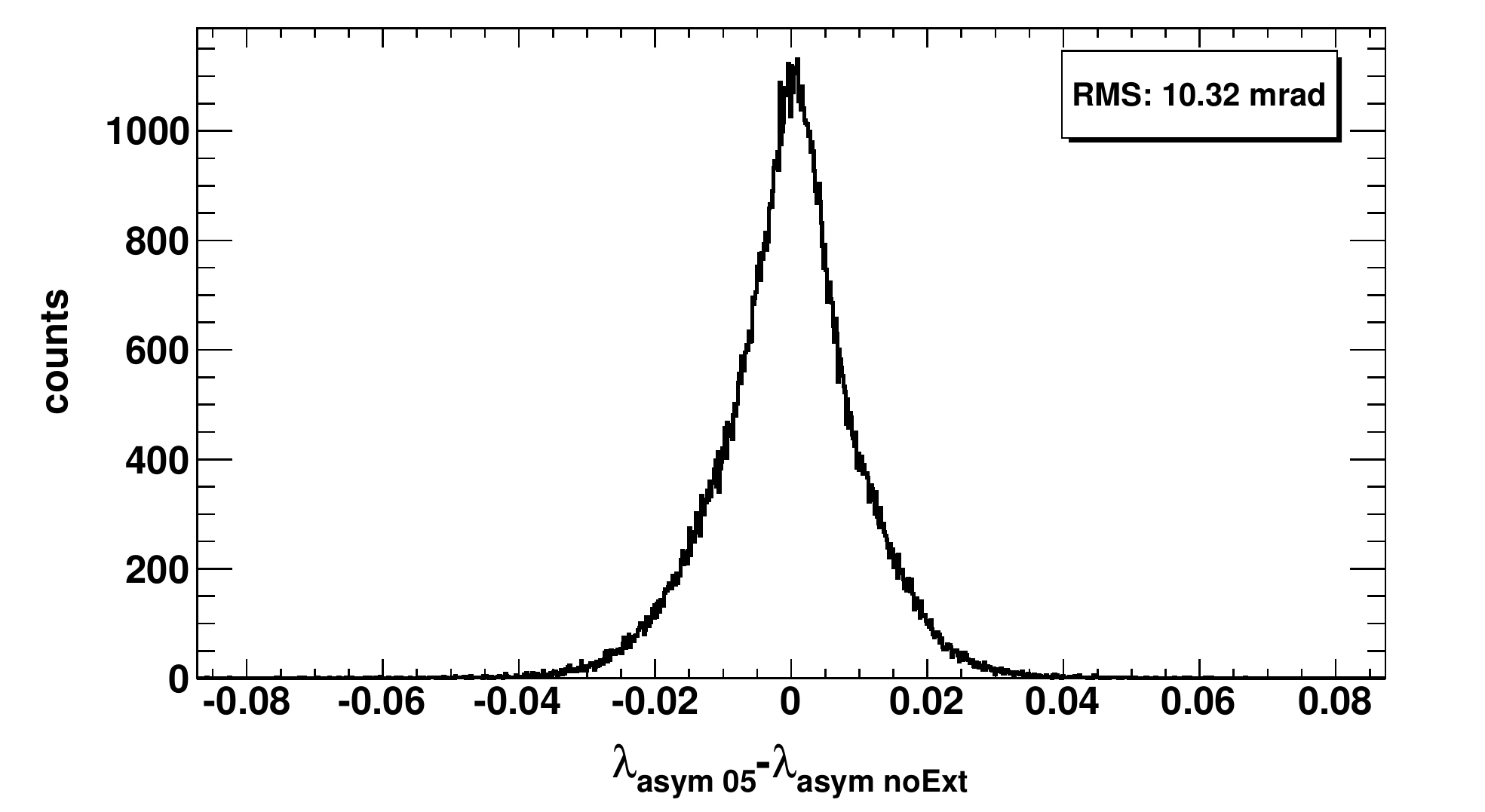}}
\hfill
\subfigure[Longitude]{\includegraphics[width=0.235\textwidth]{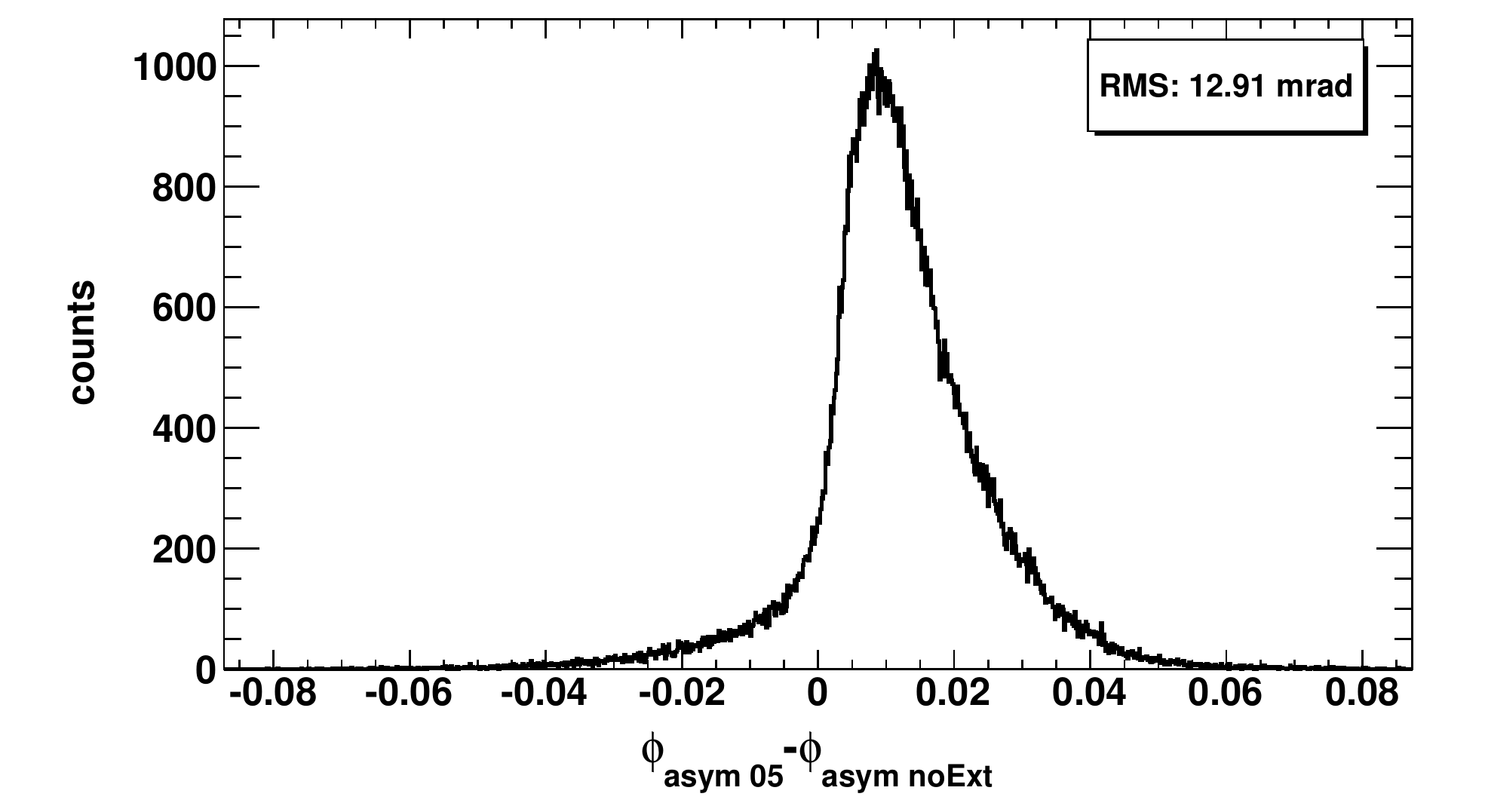}}
\hfill
\caption{Latitudinal and Longitudinal difference of last point (magnetopause) for AMS-02 electrons in the rigidity range between 20 and
  30 GV, for just internal field IGRF and T05}
\label{simp_fig1}
\end{figure}

\begin{figure}[ht!]
\hfill
\subfigure[Latitude]{\includegraphics[width=0.235\textwidth]{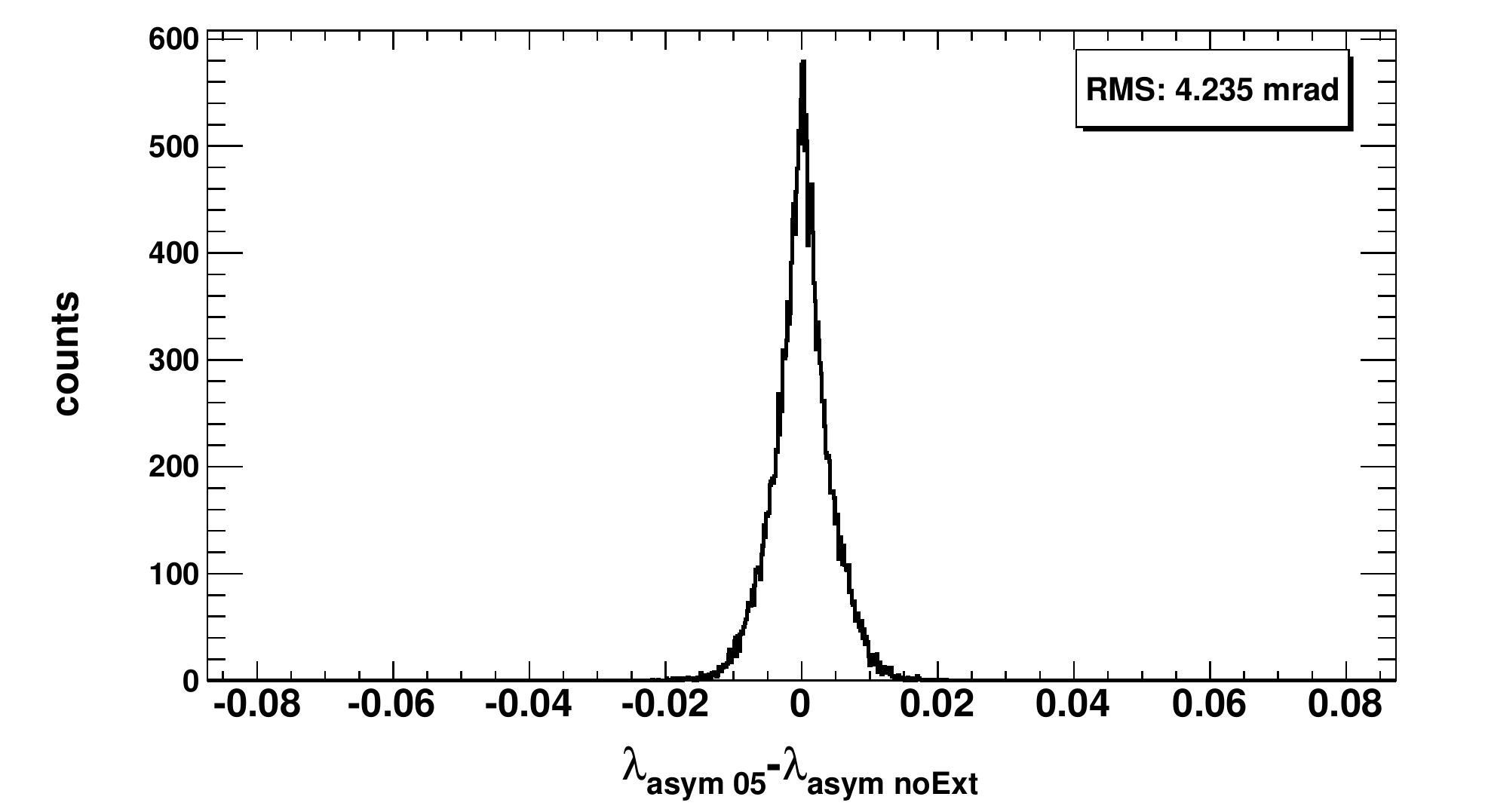}}
\hfill
\subfigure[Longitude]{\includegraphics[width=0.235\textwidth]{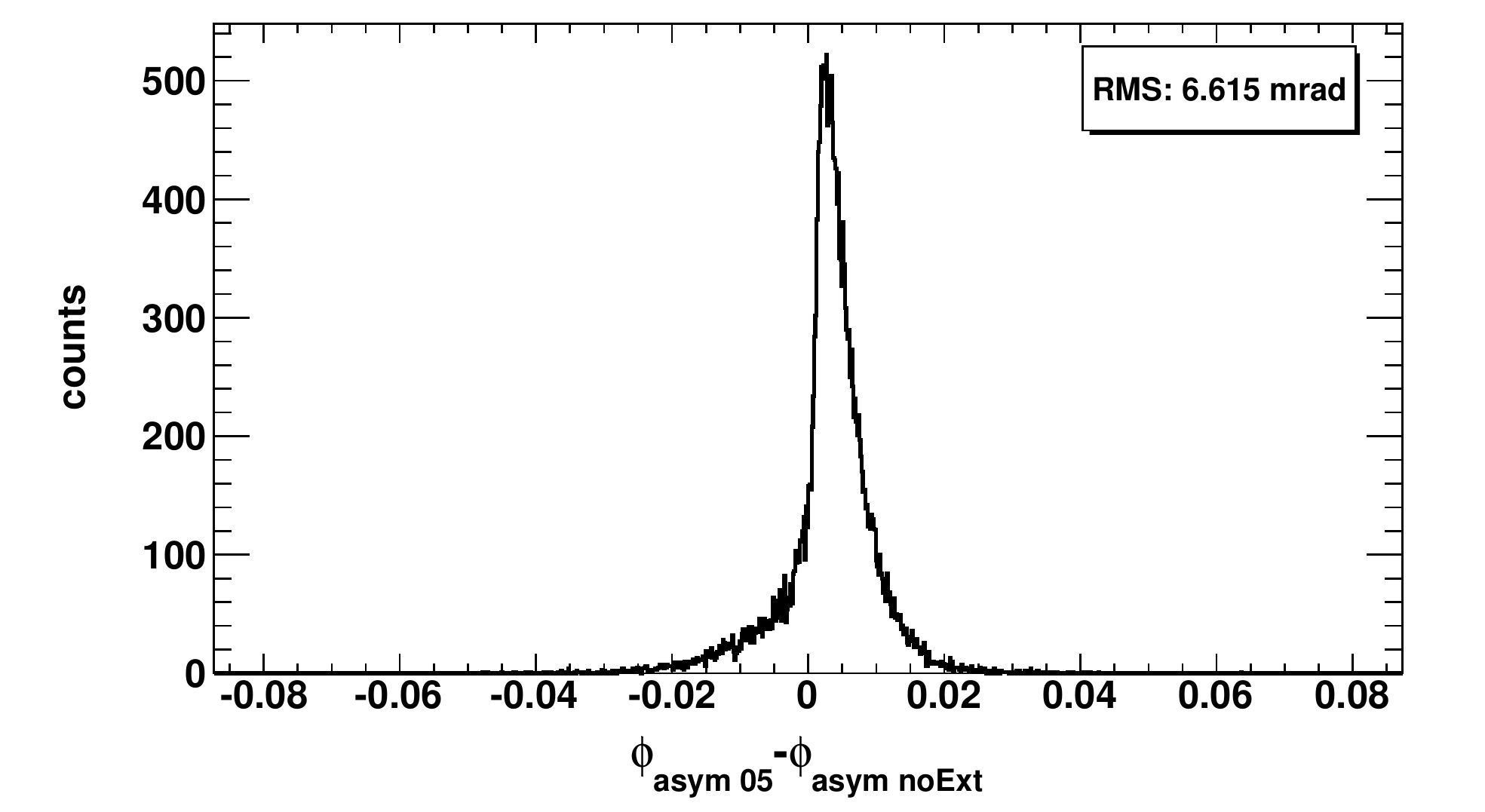}}
\hfill
  \caption{Latitudinal and Longitudinal difference of last point (magnetopause) for AMS-02 electrons for rigidity above
  50 GV, for just internal field IGRF and T05}
  \label{simp_fig2}
\end{figure}

\begin{figure}[ht!]
\hfill
\subfigure[Latitude]{\includegraphics[width=0.235\textwidth]{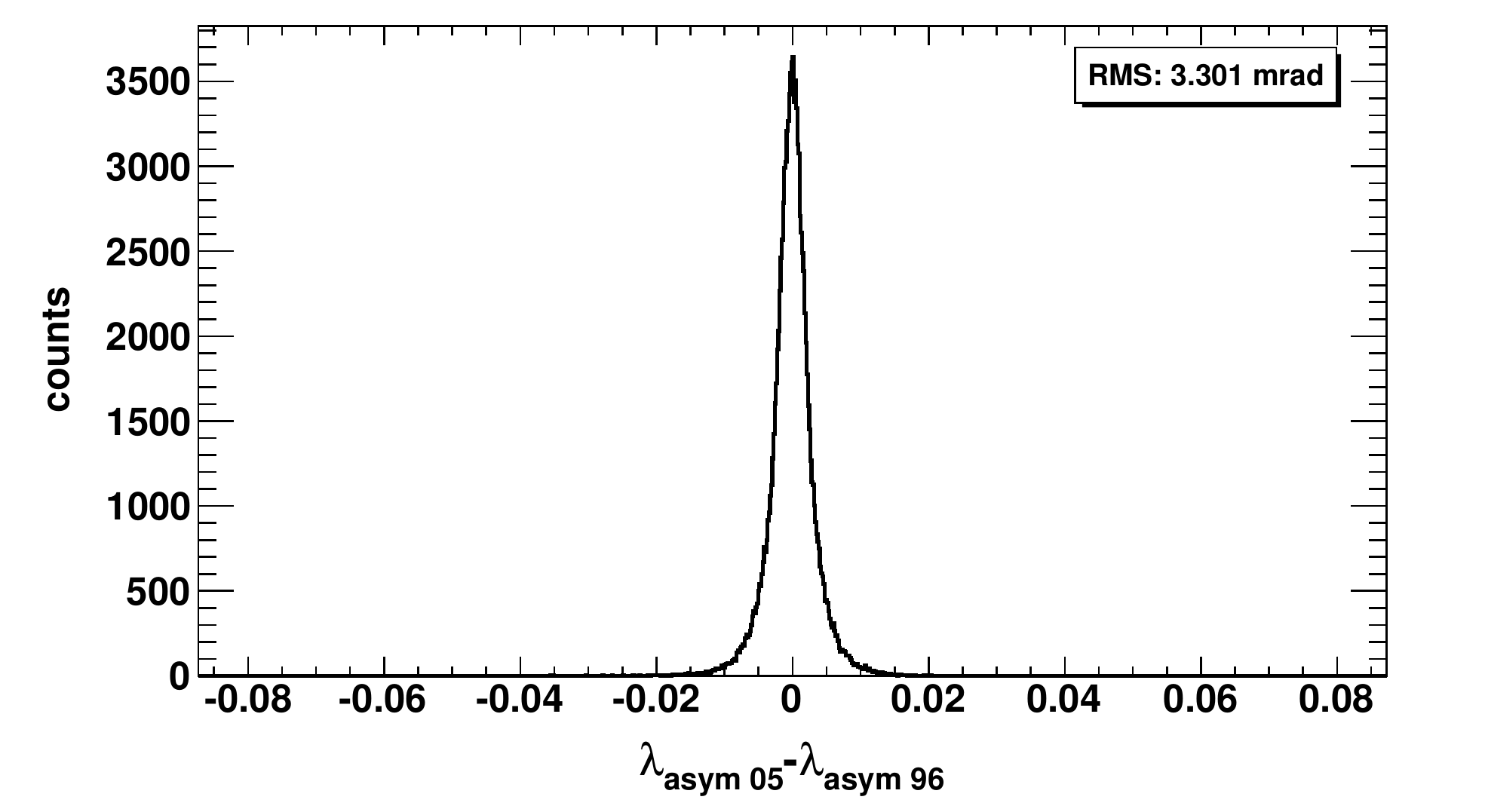}}
 \hfill
\subfigure[Longitude]{\includegraphics[width=0.235\textwidth]{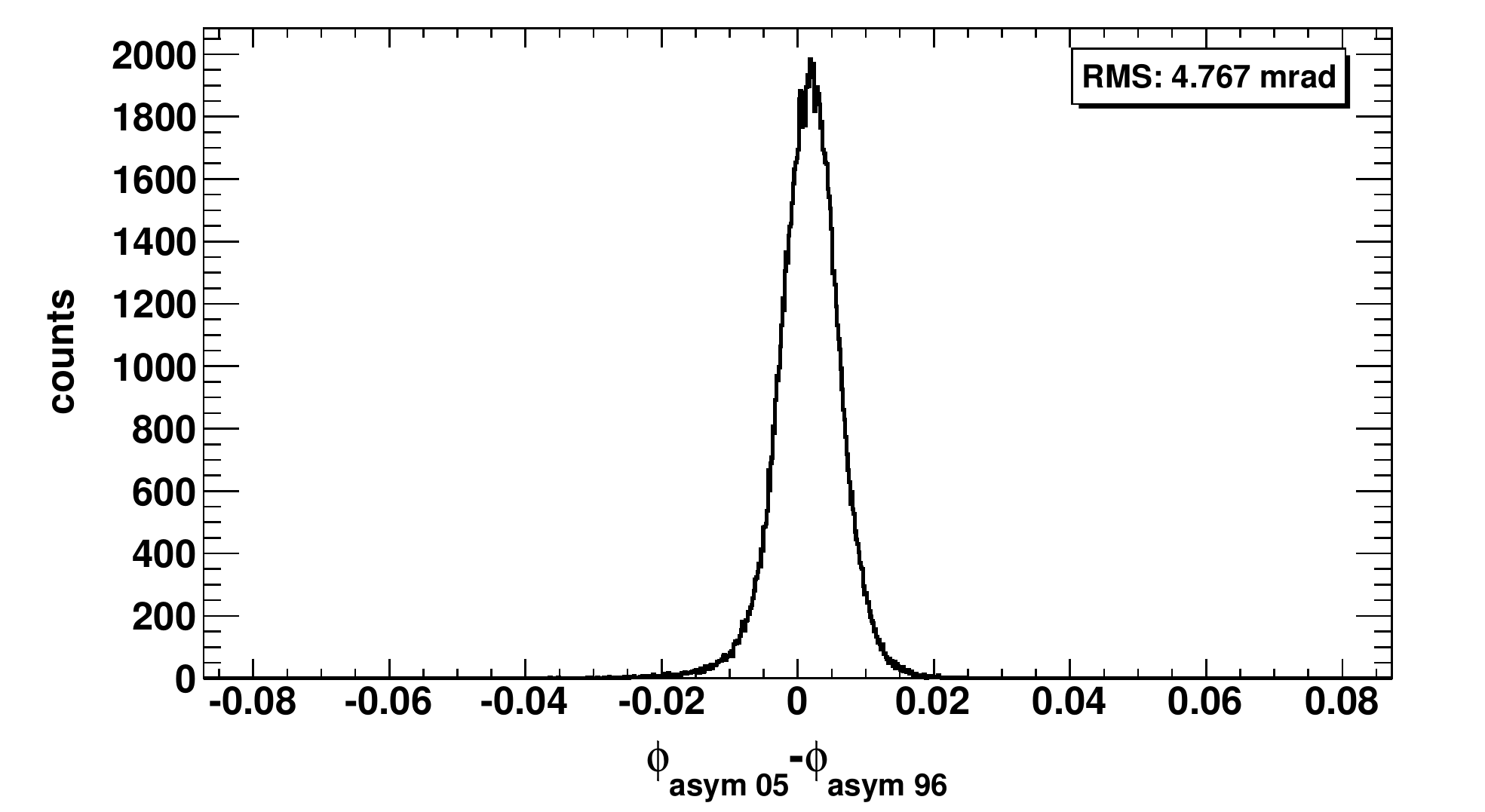}}
\hfill  
  \caption{Latitudinal and Longitudinal difference of last point (magnetopause) for AMS-02 electrons in the rigidity range between 20 and
  30 GV, for external field models T96 and T05}
  \label{simp_fig3}
\end{figure}

\begin{figure}[ht!]
\hfill
\subfigure[Latitude]{\includegraphics[width=0.235\textwidth]{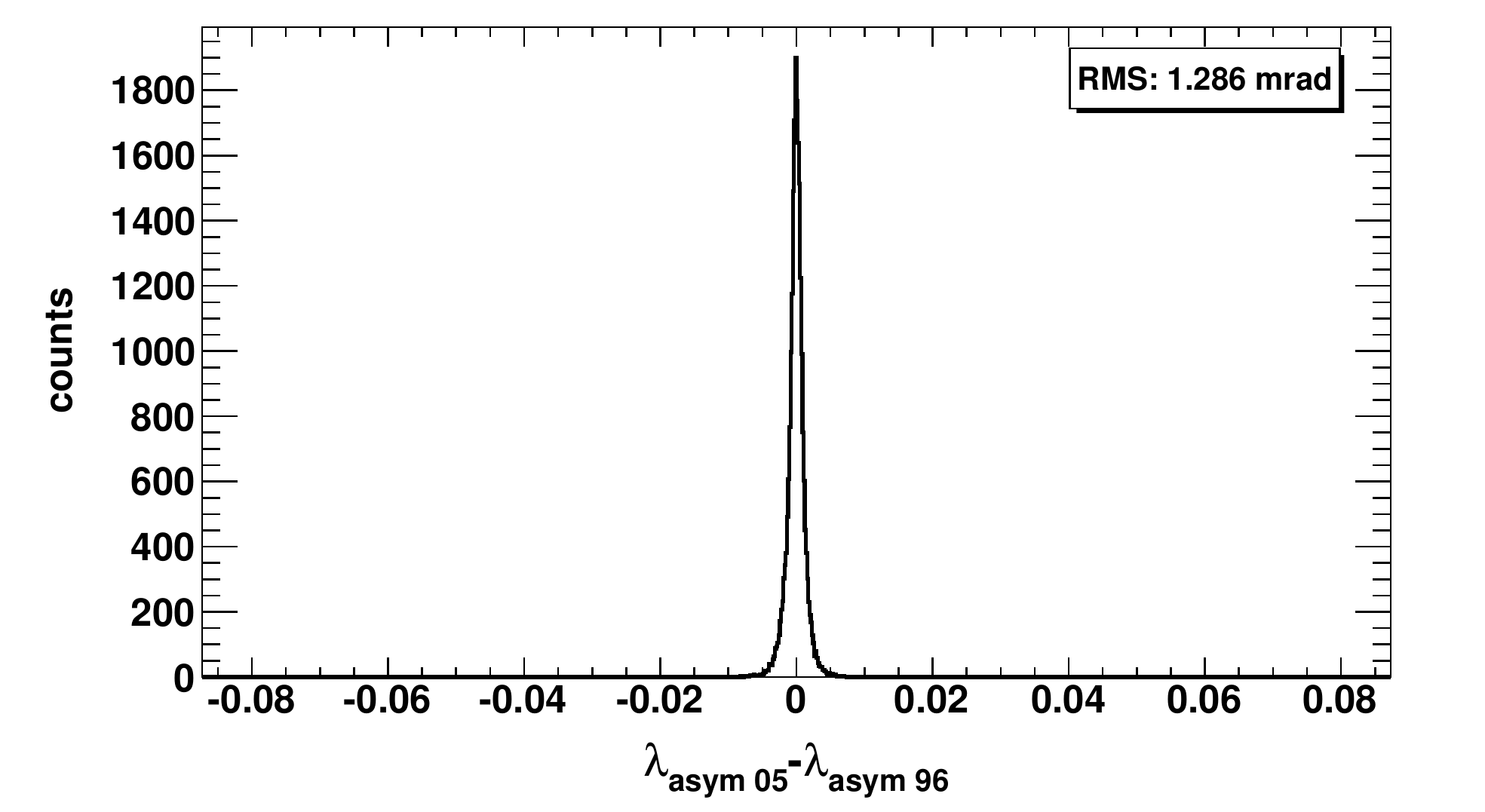}}
\hfill
\subfigure[Longitude]{\includegraphics[width=0.235\textwidth]{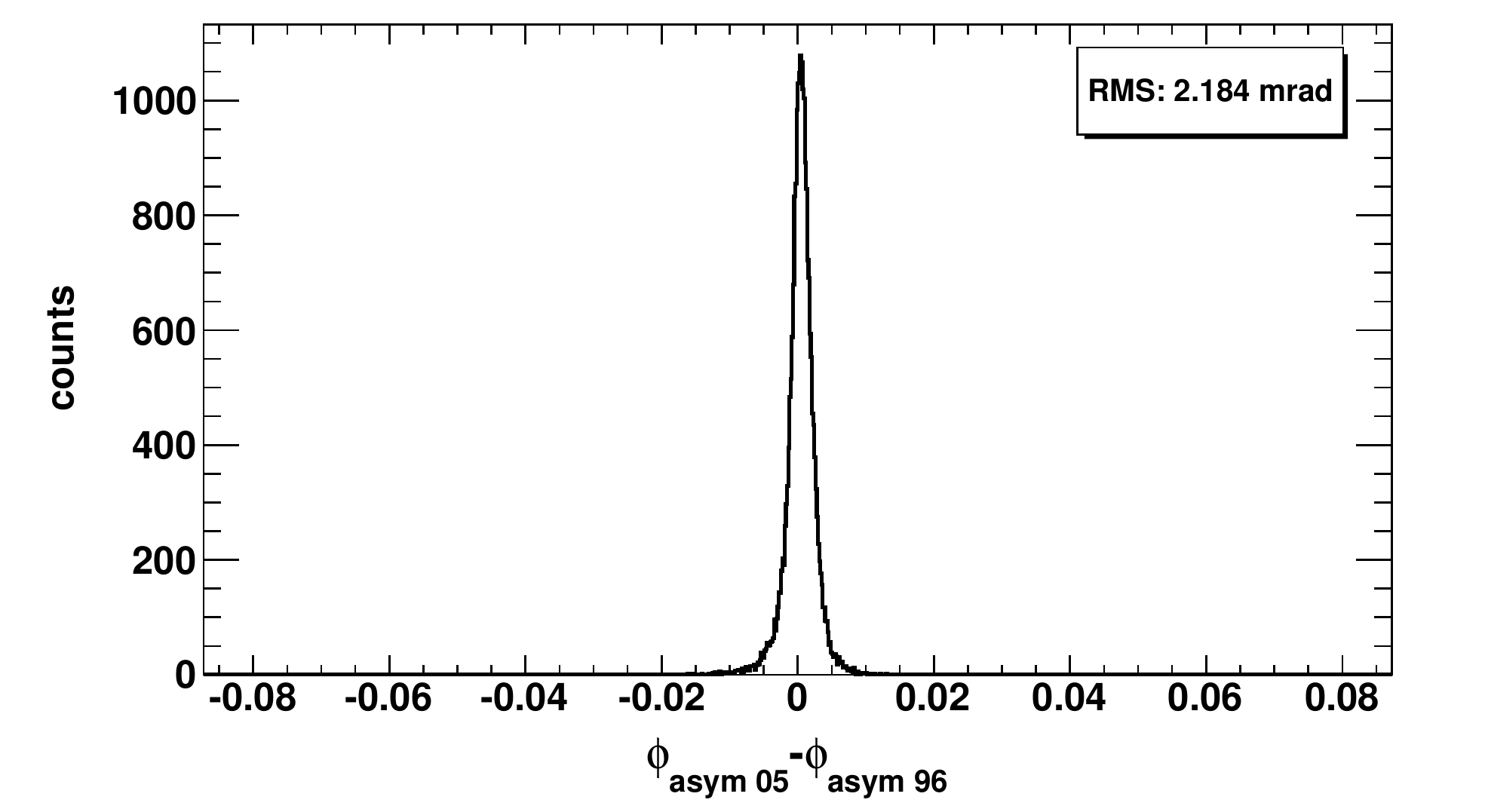}}
\hfill
  \caption{Latitudinal and Longitudinal difference of last point (magnetopause) for AMS-02 electrons for rigidity above
  50 GV, for external field models T96 and T05}
  \label{simp_fig4}
\end{figure}

Then we restricted our analysis only on primary CR, so trajectories reconstructed by both models as allowed. In this case
the main question is how precise can be the reconstructed trajectory up to the border of
the magnetopause. This question is related also to the magnetopause model and the mathematical precision of the code. As described
elsewhere (\cite{bib:ams_mib}) our backtracing code, also in his ``refurbished'' version that has been included in the AMS official
software, reconstructs a particle trajectory with an accuracy $\le 10^{-3}$ (this is an upper limit occurring in particular cases, most of the times the
precision is much better than this value). 
The evaluation of this accuracy takes into account all the parameters of the backtracing (time,
altitude, direction), and is also related to the step size choosed. A fine tuning of each source of uncertainty, as described here \cite{bib:ams_mib},
led to an optimization of the code, but every estimate will always have its intrinsic ``error'', in relation to the fact that we reproduce a curved trajectory
as a sum of infinitely small linear segments.
Once everything is fixed (epecially the magnetosphere model that has
been choosed to be for both external field models the Shue one \cite{bib:shue}) we evaluated the difference in final points for different
rigidity bins (roughly speaking the curvature due to the two different magnetic field model).
Our analysis was done on a sample of $2.2\times 10^{5}$ electrons detected by AMS-02 in the period between June 2011 and September 2012. 
Primary particles has been divided in 4 rigidity bins: 20-30 GV, 30-40 GV, 40-50 GV and $>$50 GV (here we present only first and last bin). 
First of all we show in Fig. \ref{simp_fig1} and \ref{simp_fig2} the difference, separated in Latitude and Longitude  
between only IGRF internal field and External field T05. As can be clearly seen increasing the energy (in this case rigidity) the 
difference decreases. If we compare these plots with \ref{simp_fig3} and \ref{simp_fig4}, where again are evaluated 
differences in the last points, but this time between the 2 external field models, we can see that not only it is reduced at high
rigidity, but also in the bin 20-30 GV. This is an additional indication that the external field models are essential for this kind of study, and that
they are also consistent.

Our study was focused on allowed trajectories and on the different entry point as reconstructed by backtracing 
(important in every case of direction study as for example anisotropy). 
As can be seen in Table \ref{table_diffnoext2a} and \ref{table_diffnoext2b} where we present the \% of particles whose last point difference
is greater than $0.5^{\circ}$, there is a significant difference between with and without the External field model. Moreover in 
Table \ref{table_diffnoext1a} and \ref{table_diffnoext1b} is also shown a reduction of the rms distribution both for Latitudinal and Longitudinal
difference (higher this second one due to the ``dipolar'' structure of the geomagnetic field).

\begin{table}[ht!]
\begin{center}
\begin{tabular}{|c|c|c|c|}
\hline  & All & P$<$4nPa & P$>$4nPa \\ \hline
Latititude NoExt-T05  & 10.9 & 10.3 & 18.2\\ \hline
Longitude  NoExt-T05 & 13.6 & 12.9  & 21.6 \\ \hline
Latititude T96-T05  & 3.5 & 3.3 & 6.7 \\ \hline
Longitude  T96-T05 & 5.0 & 4.5 & 8.8 \\ \hline
\end{tabular}
\caption{RMS of difference distribution in mrad - sample NoExt-T05 and T96-T05 - Bin 20-30 GV}
\label{table_diffnoext1a}
\end{center}
\end{table}

\begin{table}[ht!]
\begin{center}
\begin{tabular}{|c|c|c|c|}
\hline  & All & P$<$4nPa & P$>$4nPa \\ \hline
  $> 0.5^{\circ}$  NoExt-T05 & 78.8\% & 78.4\% & 85.1\%\\ \hline
  $> 0.5^{\circ}$ T96-T05  & 10.6\% & 9.4\% & 38.8\%\\ \hline
\end{tabular}
\caption{\% of particles with difference in last point greater than $ 0.5^{\circ}$ - NoExt-T05 and T96-T05 -  Bin 20-30 GV}
\label{table_diffnoext1b}
\end{center}
\end{table}

\begin{table}[ht!]
\begin{center}
\begin{tabular}{|c|c|c|c|}
\hline  & All & P$<$4nPa & P$>$4nPa \\ \hline
Latititude NoExt-T05  & 4.5 & 4.3 & 7.1\\ \hline
Longitude  NoExt-T05 & 6.9 & 6.7  & 10.6 \\ \hline
Latititude T96-T05  & 1.5 & 1.4 & 2.6 \\ \hline
Longitude  T96-T05 & 2.3 & 2.2 & 3.7 \\ \hline
\end{tabular}
\caption{RMS of difference distribution (in mrad) - sample NoExt-T05 and T96-T05 - Bin $>$50 GV}
\label{table_diffnoext2a}
\end{center}
\end{table}

\begin{table}[ht!]
\begin{center}
\begin{tabular}{|c|c|c|c|}
\hline  & All & P$<$4nPa & P$>$4nPa \\ \hline
  $> 0.5^{\circ}$  NoExt-T05 & 23.4\% & 22.1\% & 47.9\%\\ \hline
  $> 0.5^{\circ}$ T96-T05  & 0.8\% & 0.6\% & 5.0\%\\ \hline
\end{tabular}
\caption{Percentage of particles with difference in last point greater than $ 0.5^{\circ}$ - NoExt-T05 and T96-T05 -  Bin $>$50 GV}
\label{table_diffnoext2b}
\end{center}
\end{table}

\subsection{Tsyganenko 96 or 2005}
Our study was then divided in two main groups, in relation to the Solar Wind pressure (the main indication of magnetic disturbancies due to solar
activity). This was one mainly because, as can be seen in Table \ref{table_missing}, parameters for T05, recalculated as specified in 
Sec. \ref{subsec:int_ext} from measured parameters and, also in they are interpolaed in case of ``holes'', often data are missing and parameters can
not be obtained (see \cite{bib:05REP} for available parameters from Tsyganenko evaluation). Due to the fact that below 4 nPa (2 times the average
value of Solar Wind pressure) the difference between the 2 External Field models is lower, we evaluated the possibility to avoid such a loss o information 
($\simeq$ 15\%) using T96 model during ``quiet'' periods. The difference between last points for the two different models is indeed greater for high 
(i.e $P>4~nPa$) solar activity period, as can be seen from rms in Table \ref{table_missing}. 
The choice to use mainly T05 model for AMS-02 data analysis is related to the fact that we would like to test it in this high solar activity period,
as it has been developed especially for storm events, that as reported also in our website (see \cite{bib:geomagsphere}), are quite frequent. 
\begin{table}[ht!]
\begin{center}
\begin{tabular}{|c|c|c|c|}
\hline  & T96  & T05 & T96 no T05\\ \hline
  \% missing  & 1.5\% & 14.8\% &\\ \hline
  \% fill  & &  & 13.3\%\\ \hline
  \% fill ($<$4 nPa)  & &  & 12.6\% \\ \hline
\end{tabular}
\caption{Missing parameters for T96 and T05 models, and \% of recovered information using T96 below 4 nPa}
\label{table_missing}
\end{center}
\end{table}
As can be seen in Table \ref{table_missing}, using T96 when Pdyn $<$ 4 nPa we can reduce missing parameters below 2\% of the whole period.

\section{Conclusions}
We developed a code to reconstruct charge particle trajectory inside the geomagnetic field. This code has been implemented in the official software of
the AMS-02 experiment in order to be used to separate primary and secondary CR. We evaluated the need for any kind of study (especially if related to
the arrival direction of particles as for example anisotropy) to include the external field model, and we warmly recommend to use the last 
Tsyganenko 2005 model, developed for storm periods like that one in which AMS-02 is taking data. We applied our code to primary electron selected in
more than one year of data and we suggested the possibility to use the Tsyganenko 1996 model to fill missing parameters for T05, when the Soalr Wind
pressure is below a certain limit, choosen as 2 times the average value.

\section{Acknowledgements}
We wish to especially thank Prof. N. Tsyganenko that provided us the new parameters for T05 model. 
This work is supported by Agenzia Spaziale Italiana under contract ASI-INFN I/002/13/0, Progetto AMS - Missione scientifica ed analisi
dati.

\vspace*{0.5cm}

\end{document}